\documentstyle[aps,prb,floats,epsf]{revtex}

\begin{document}
\draft


\twocolumn[\hsize\textwidth\columnwidth\hsize\csname@twocolumnfalse%
\endcsname
\title{Orbital dynamics: The origin of the anomalous 
optical spectra in ferromagnetic manganites}
\author{R.\ Kilian and G.\ Khaliullin \cite{KHA}}
\address{Max-Planck-Institut f\"ur Physik komplexer Systeme,
N\"othnitzer Strasse 38, D-01187 Dresden, Germany}
\date{June 30, 1998}
\maketitle


\begin{abstract}
We discuss the role of orbital degeneracy in the transport 
properties of perovskite manganites, focusing in
particular on the optical conductivity in the metallic
ferromagnetic phase at low temperatures. Orbital degeneracy
and strong correlations are described by an orbital 
$t$-$J$ model which we treat in a slave-boson approach.
Employing the memory-function formalism we calculate the 
optical conductivity, which is found to exhibit a broad 
incoherent component extending up to bare bandwidth
accompanied by a strong suppression of the Drude weight.
Further, we calculate the constant of $T$-linear
specific heat. Our results are in overall agreement with
experiment and suggest low-energy orbital fluctuations
as the origin of the strongly correlated nature of the metallic
phase of manganites.
\end{abstract}

]


Perovskite manganese oxides, $R_{1-x}A_x$MnO$_3$ (where $R$ and $A$ 
represent rare-earth and divalent metal ions, 
respectively) exhibit rich physical behavior \cite{RAM97} 
whose origin lies in the mutual interplay between spin, 
charge, orbital, and lattice degrees of freedom as well as in the
strong correlations among electrons. The key elements of electronic 
structure are the Mn $3d$ orbitals, which due to crystal-field 
splitting and strong Hund coupling are in a $(t_{2g})^3(e_g)^1$ 
configuration for $x=0$, with holes being induced in the twofold 
degenerate $e_g$ orbitals for $x>0$. 
Electrons in the $t_{2g}$ and $e_g$ orbitals are usually
considered as localized and itinerant, respectively.

Early work on the transport properties of manganites has focused 
primarily on the double-exchange scenario, \cite{ZAG} which also,
competing with antiferromagnetism, results in a complex
magnetic phase diagram. The renewed interest in manganese 
oxides sparked by the recent discovery of a very large magnetoresistance
has led to a controversial discussion of the role of lattice and orbital 
degrees of freedom; the effect of onsite correlations among $e_g$ electrons, 
however, has mostly been discarded.
On the experimental side, these correlations seem to be strong even
in the metallic state of manganites, despite the fact that double exchange 
strongly suppresses the spin dynamics below $T_c$. This can be deduced, e.g.,
from the observation of only a small discontinuity at the
Fermi level in photoemission spectra. \cite{CHA93,SAI95,SAR96} Particularly 
striking in this respect are recent measurements of the optical
conductivity of La$_{1-x}$Sr$_x$MnO$_3$. \cite{OKI95} The conductivity 
in the metallic state was found to consist of two components: 
(1) a narrow Drude peak at $\omega<0.02\text{ eV}$ with strongly 
suppressed weight and (2) a broad incoherent part extending
up to $\omega \approx 1\text{ eV}$. It is remarkable that the
incoherent component remains finite even at $T \ll T_c$ where
ferromagnetic moments have completely saturated, clearly
indicating the presence of other degrees of freedom besides
spin fluctuations. The fact that the incoherent part of the
spectrum extends to rather high frequencies suggests 
these degrees of freedom to be of electronic origin,
possibly resulting from the orbital degeneracy of $e_g$ orbitals
as has been proposed by several authors. \cite{SHI97,ISH97}

Shiba, Shiina, and Takahashi \cite{SHI97} ascribe the excitations leading to 
the incoherent structure of the optical conductivity to transitions between 
two fermionic bands within the degenerate $e_g$ orbitals, which
they find to result in a spectrum extending up to
fermionic bandwidth. While their model indicates the
role of orbital degeneracy, it neglects the $e_g$-electron
correlations. A more elaborate treatment of both, orbital
degeneracy and electron correlations, was done by Ishihara, 
Yamanaka, and Nagaosa. \cite{ISH97} Starting from an orbital-charge separation
scheme these authors showed that orbital fluctuations are strong enough
to prevent orbital ordering from developing.
Performing numerical calculations based on a phenomenological model
that simulates orbital disorder by static randomness in the hopping 
matrix elements they further were able to recover the characteristic features 
of the optical conductivity spectrum, though no Drude
component was obtained. 

In this paper we report on the first
microscopic theory of the optical conductivity which combines 
strong correlations and orbital degeneracy. The transport properties
of manganites are shown to be highly incoherent even in the
ferromagnetic phase due to strong scattering of charge
carriers on dynamical orbital fluctuations.  This gives rise to a broad 
optical absorption spectrum extending up to bare bandwidth accompanied by 
a strong suppression of the Drude weight. The theory further accounts for
the small values of specific heat found experimentally. \cite{WOO97}

Assuming complete ferromagnetic saturation of electronic spins, 
which will henceforth be neglected, 
our starting point is an orbital $t$-$J$ model:
\begin{eqnarray}
H &=& -\sum_{\langle ij \rangle_{\gamma}} \sum_{\alpha\beta} 
\left(t_{\gamma}^{\alpha \beta}
\hat{c}^{\dagger}_{i\alpha} \hat{c}_{j\beta}+\text{H.c.}\right)\nonumber\\
&& + J \sum_{\langle ij \rangle_{\gamma}}\left( \tau_i^{\gamma}
\tau_j^{\gamma} - \frac{1}{4} n_i n_j \right), 
\label{HTJ}
\end{eqnarray}
where $\langle ij \rangle_{\gamma}$
indicates summation over manganese nearest-neighbor bonds in spatial
direction 
$\gamma \in \left\{\hat{x},\hat{y},\hat{z}\right\}$.
The operator $\hat{c}^{\dagger}_{i\alpha} = 
c^{\dagger}_{i\alpha}\left(1-n_i\right)$ creates an electron
at an empty site $i$ with orbital pseudospin $\alpha$ for 
which we us the notation
$\uparrow = d_{3z^2-r^2}$ and
$\downarrow = d_{x^2-y^2}$.
The anisotropic transfer matrix elements are given by \cite{KUG73}
\[
t_{x/y}^{\alpha\beta} = 
t\left(\begin{array}{cc}
1/4 & \mp\sqrt{3}/4\\
\mp\sqrt{3}/4 & 3/4
\end{array}\right), \quad
t_{z}^{\alpha\beta} = 
t\left(\begin{array}{cc}
1 & 0\\
0 & 0
\end{array}\right),
\]
allowing for interorbit transitions in the $xy$ plane.
The strength of the pseudospin interaction is controlled by the 
superexchange coupling constant $J=2t^2/U$, where $U$ 
is the onsite repulsion between spin-parallel $e_g$ 
electrons. \cite{ISH96} Since $U$ is believed to be 
large we assume $J < t$. 
The pseudospin operators of Eq.~(\ref{HTJ}) are defined as
\[
\tau^{x/y} = \frac{1}{4}\left(\sigma^z\pm
\sqrt{3}\sigma^x\right),\quad
\tau^{z} = \frac{1}{2}\sigma^z,
\]
with Pauli matrices $\sigma^x$ and $\sigma^z$. 
We note that in the real system Jahn-Teller coupling further
contributes to the $J$ term of Hamiltonian~(\ref{HTJ}).
Three-site hopping terms that are contained in a general
orbital $t$-$J$ model \cite{HOR98} are neglected for simplicity. 

We calculate the optical conductivity $\sigma(\omega)$ using the
memory-function formalism. \cite{GOE72} While not rigorous this
method yields exactly the leading terms of a high-frequency 
expansion of $\sigma(\omega)$ and is believed to give reasonably
accurate results over the whole frequency range if no critical
low-energy modes as in one-dimensional systems exist. Within
this framework the optical conductivity $\sigma(\omega)$ is expressed 
via the memory function $M(\omega)$ 
\begin{equation}
\sigma(\omega) = \chi_0 \frac{i}{\omega+M(\omega)},
\label{SIG}
\end{equation}
where $\chi_0$ is the zero-frequency current-current correlation
function. The memory function is given by
\begin{equation}
M(\omega) = -\frac{1}{\omega \chi_0} \big[f(\omega)-f(0)\big],
\label{MEM}
\end{equation}
with $f(\omega)$ being the correlation function of
force operators $F_z = \left[J_z,H\right]$. The 
transport properties of the system are isotropic in the 
orbital-disordered phase, allowing us to select
the current along the $z$ direction which is of the simple form
\begin{equation}
J_z = -\frac{iet}{Na^3}\sum_{\langle ij \rangle_z}
\left( \hat{c}^{\dagger}_{i \uparrow}
\hat{c}_{j \uparrow}-\text{H.c.}\right),
\end{equation}
where $N$ is the number of lattice sites and $a$ denotes the
lattice constant.
The force operator consists of two parts $F_z^t$ and $F_z^J$ 
corresponding to the $t$ and $J$ term of Hamiltonian~(\ref{HTJ}).
Due to the relative smallness of the coupling constant $J$ as compared 
to $t$ the latter
term is expected to contribute only little to the optical
conductivity and will therefore be neglected. For the former term
one finds
\begin{eqnarray}
F^t_z &=& -\frac{iet^2}{2Na^3} \sum_{i,\gamma,\delta}
\sum_{\alpha\beta}
\left(R^z_{i+\delta}-R^z_i\right) t_{\gamma}^{\alpha\beta}\nonumber\\ 
&&\times\left(B_{i\alpha} \hat{c}^{\dagger}_{i+\delta\uparrow}
\hat{c}_{i+\gamma,\beta}+\text{H.c.}\right),
\label{FOR}
\end{eqnarray}
with the bosonic operator
\[
B_{i\alpha} = \left(2-n_i+\sigma_i^z\right)\delta_{\alpha\uparrow}
+\left(\sigma_i^x-i\sigma_i^y\right)\delta_{\alpha\downarrow},
\]
where double occupancy of sites is excluded.

We employ a slave-boson theory to express
Hamiltonian~(\ref{HTJ}) and the force operator, Eq.~(\ref{FOR}), 
in terms of fermionic spinon operators $f_{i\alpha}$
and bosonic holon operators $h_i$. The pseudoparticles fulfill 
the local constraint
$f^{\dagger}_{i\uparrow}f_{i\uparrow} + f^{\dagger}_{i\downarrow}
f_{i\downarrow} + h^{\dagger}_i h_i = 1$ which is relaxed
to a global one. On a mean-field level, the dynamics of spinons 
and holons are given by decoupling Hamiltonian~(\ref{HTJ})
into fermionic and bosonic parts, introducing mean-field parameters 
$\chi = \langle f^{\dagger}_{i\uparrow} f_{j\uparrow} \rangle_z$ and
$\sqrt{x} = \langle h_i \rangle$, where $x$ is the concentration
of doped holes. The mean-field Hamiltonian describes two fermionic 
Gutzwiller bands of width $W_s = 6(xt+\chi J)$ and one bosonic band 
of width $W_h = 12\chi t$.

From analogy to the conventional $t$-$J$ model the
slave-boson representation with its fermionic description of
low-lying orbital excitations is expected to describe well an orbitally
disordered state far from critical instabilities towards orbital
ordering. The existence of such an orbital 
liquid state is strongly supported by the analysis of Ishihara, 
Yamanaka, and Nagaosa \cite{ISH97} who find orbital disorder even within a 
slave-fermion Schwinger-boson representation which favors orbital
ordering. Assuming the stability of the disordered state, 
we select the slave-boson representation, noting, however, that short-range 
orbital correlations will not be fully captured in this picture.

\begin{figure}[t]
\noindent
\centering
\epsfxsize=0.9\linewidth
\epsffile{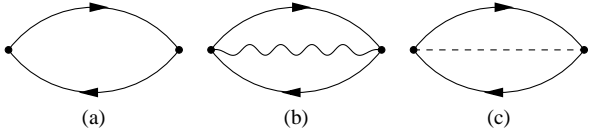}\\[3pt]
\caption{Contributions to the force-force correlation function.
Solid, wavy, and dashed lines denote spinon, holon, and pseudospin
Green's functions, respectively.}
\label{DIA}
\end{figure}

The memory function, Eq.~(\ref{MEM}), is calculated via
the force-force correlation function, replacing holon 
operators by their mean-field value and keeping fluctuations around
this value up to first order. Diagram~(a) of Fig.~\ref{DIA}
describes transitions between the two coherent Gutzwiller 
bands of the spinon mean-field Hamiltonian. These interband 
transitions are allowed due to the fact that the $t$ term of 
Hamiltonian~(\ref{HTJ}) is not diagonal in orbital quantum 
numbers. Using a free-fermion picture, similar transitions were 
found by Shiba, Shiina, and Takahashi \cite{SHI97} to contribute to the optical 
conductivity spectrum. The presence of electron correlations, however, 
strongly suppresses the spectral weight associated with these processes
and shifts the upper bound of the corresponding
absorption spectrum to low energies. 

Diagrams~(b) and (c) of Fig.~\ref{DIA} purely originate from
correlations among electrons and are absent in a free theory. They 
describe transitions between a coherent Gutzwiller and a 
highly-incoherent band 
with corresponding contributions to the optical conductivity 
spectrum that extend approximately up to bare fermionic bandwidth.
We note that such broad incoherent absorption was also found in a 
recent finite-temperature diagonalization study performed on 
small clusters. \cite{HOR98} 
Diagram~(b) accounts for the composite nature of 
charge carriers in a strongly correlated system reflected
by the spinon-holon convolution. The force operator employed in the 
calculation of this diagram yields terms involving two as well as three 
lattice sites. The two-site terms are already contained in the 
current operator and thus also appear in a direct calculation 
of the optical conductivity via the current-current correlation
function. The three-site terms, however, are intrinsic to the force-force 
correlation function and partially account for vertex corrections.
Diagram~(c) finally describes scattering of carriers on 
pseudospin fluctuations. We note that the force term $F_z^J$ neglected 
before contains only terms involving three lattice sites of small 
amplitude, supporting our previous approximation to discard this term. 

\begin{figure}[t]
\noindent
\centering
\epsfxsize=0.9\linewidth
\epsffile{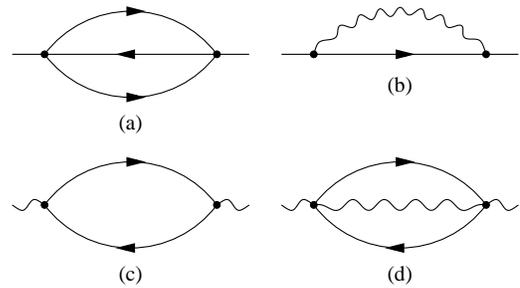}\\[0pt]
\caption{Self-energies of (a), (b) spinons and (c), (d) holons.}
\label{SEL}
\end{figure}

The spinon and holon propagators of Fig.~\ref{DIA} are
calculated with the self-energy corrections shown in 
Fig.~\ref{SEL}. The spinon self-energy diagrams~(a) and (b)
describe scattering on pseudospin and holon-density fluctuations, 
respectively. The quasiparticle weight, which is proportional to $x$ 
on the mean-field level, is further suppressed by these
incoherent scattering processes. This smallness of quasiparticle
weight is consistent with the experimental observation of
only a small discontinuity in the spectral-weight function at the
Fermi energy as seen in photoemission spectra. \cite{CHA93,SAI95,SAR96}
The bosonic self-energy diagrams~(c) and (d) describe scattering on
fluctuations of the spinon bond-order parameter, creating a broad 
incoherent peak in the bosonic spectral function.

The expressions obtained from the diagrams in Figs.~\ref{DIA}
and \ref{SEL} contain several integrations over momentum
space which we solve numerically using a Monte Carlo algorithm.
The zero-frequency current-current
correlation function $\chi_0$ of Eqs.~(\ref{SIG}) and (\ref{MEM}) 
is obtained from the spinon mean-field Hamiltonian 
as $\chi_0 = 2x\chi t e^2/a$.
We choose a hole-doping concentration of $x=17.5\%$ and set
$J=0.4t$. The value of the spinon bond-order parameter is 
numerically determined to be $\chi=0.25$,
the lattice constant is set to $a=3.9\text{ \AA}$.
The real part of the optical conductivity, Eq.~(\ref{SIG}), 
consists of the Drude component $\sigma_D(\omega) = \pi \chi_0 
\big[1+\Re(dM/d\omega)|_{\omega\rightarrow 0}\big]^{-1}
\delta(\omega)$ and the regular part 
$\sigma_{\text{reg}}(\omega)$. The spectrum 
of the regular part is shown in Fig.~\ref{RES} indicated by
a solid line. We 
compare our theoretical result to experimental data obtained by Okimoto 
{\it et al.} \cite{OKI95} for $17.5\%$ doped La$_{1-x}$Sr$_x$MnO$_3$ at 
$T=9\text{ K}$ represented by the dashed line in Fig.~{\ref{RES}.
The only free parameter which we use to fit the theoretical to the
experimental curve is the free fermionic bandwidth which we fix by setting 
$t=0.36\text{ eV}$ consistent with band structure calculations. \cite{PIC96}

\begin{figure}[t]
\noindent
\centering
\epsfxsize=\linewidth
\epsffile{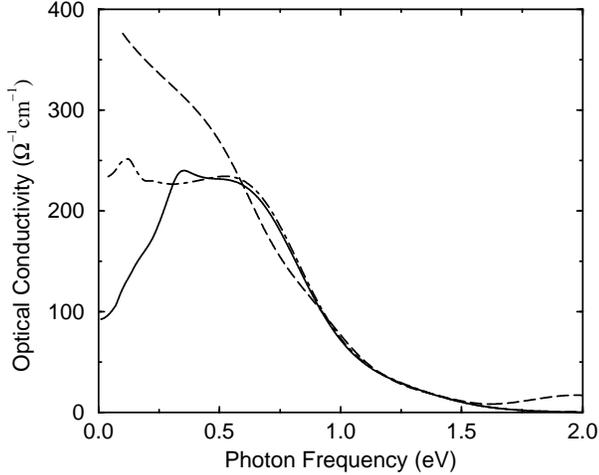}
\caption{Incoherent part of the optical conductivity 
$\sigma_{\text{reg}}(\omega)$
as a function of photon frequency $\omega$.
The solid line represents the theoretical curve for
$x=17.5\%$, which is fitted to experimental data (dashed
line) obtained by Okimoto {\it et al.} \protect\cite{OKI95}
for  La$_{1-x}$Sr$_x$MnO$_3$ at $T=9\text{ K}$. The dot-dashed
line corresponds to the theory including lattice effects.}
\label{RES}
\end{figure}

Good agreement between experiment and theory is found for intermediate and 
high frequencies. \cite{COM1} We stress that 
no additional fitting parameter is needed to obtain 
the correct absolute values of $\sigma_{\text{reg}}(\omega)$. 
The total spectral weight consisting of the Drude part $\sigma_D(\omega)$
and the incoherent part $\sigma_{\text{reg}}(\omega)$ 
is in agreement with experiment as can be seen from the 
effective charge-carrier concentration defined as
\begin{equation}
N_{\text{eff}} = \frac{2m_0a^3}{\pi e^2} 
\int_0^{\infty} d\omega \big(
\sigma_D(\omega) + \sigma_{\text{reg}}(\omega)\big);
\end{equation}
theory and experiment \cite{OKI95} both yield 
$N_{\text{eff}} = 6\%$. 

In the low-frequency region the theoretical curve deviates from
the experimental one as a pseudogap opens in the spectrum. 
As a result our theory does not completely account for 
the weight transferred to the incoherent part of the spectrum as is
reflected by the ratio of total spectral weight to Drude weight; 
here we obtain a value of $2.8$ which compares to $\approx 5$ found
experimentally. \cite{OKI95}

The discrepancy between theory and experiment in the low-frequency
part of the optical absorption spectrum indicates an 
additional scattering mechanism to be active at low energies
which is not incorporated in our theory. 
We speculate this mechanism to stem from the closeness
of the real system to orbital ordering underestimated
in the above slave-boson treatment. Scattering on
low-lying orbital collective modes induced by 
electronic superexchange and Jahn-Teller coupling 
as well as scattering on phonons will
enhance the low-energy region of the spectrum, 
thus filling the pseudogap. 
To make the latter point more explicit we calculate the
conductivity including electron-phonon effects. Coupling
of $e_g$ pseudospins to double-degenerate Jahn-Teller phonons
and of charge to the lattice breathing mode are described by
\begin{eqnarray}
H_{\text{JT}} &=& -\lambda_{\text{JT}}
\sum_i\Big(
\big(a_i^{\dagger}+a_i\big) \sigma_i^z
+i\big(a_i^{\dagger}-a_i\big)\sigma_i^x\Big),\\
H_{\text{ch}} &=& -\lambda_{\text{ch}}
\sum_i
\big( b_i^{\dagger}+b_i\big) \big(1-n_i\big),
\end{eqnarray}
respectively. For simplicity, phonons are considered to have
dispersionless energy, which we set to $\omega_0=0.05\text{ eV}$. 
Corrections to the force-force correlation function and to the fermionic 
self-energies are evaluated within a weak-coupling scheme,
assuming the dimensionless coupling constant 
$\zeta=\lambda^2 N(\epsilon_F)/\omega_0$, where 
$N(\epsilon_F)$ is the total $e_g$-density of states at the 
Fermi level, to be small below $T_c$. The result, which for 
$\zeta_{\text{JT}}=\zeta_{\text{ch}}=0.3$
is shown by the dot-dashed line in Fig.~\ref{RES}, suggests that 
(possibly strong) lattice effects are present even
in the metallic ferromagnetic phase.

We further calculate the constant of $T$-linear specific heat $\gamma$
from the spinon mean-field Hamiltonian. For $x=17.5\%$ we
find $\gamma = 7.2\text{ mJ/mol K}^2$ as compared to 
experimental values $5$ - $6\text{ mJ/mol K}^2$ (Ref.~\onlinecite{OKI95})
and $3.3\text{ mJ/mol K}^2$ (Ref.~\onlinecite{WOO97}). Experimentally $\gamma$
was observed to be nearly independent of $x$. In our theory
we find $\gamma \propto t N^0(\epsilon_F)/(xt+\chi J)$; since the
bare density of states $N^0(\epsilon)$ exhibits a pseudogap centered 
around the chemical potential at half-filling, $\gamma$ is rather 
insensitive to changes in $x$ for moderate hole-doping concentrations; 
for $x=30\%$ we in fact find $\gamma = 7.1\text{ mJ/mol K}^2$.

In summary, we have calculated the optical absorption spectrum
of ferromagnetic manganites, emphasizing the role of low-energy
orbital fluctuations leading to the strongly correlated
nature of the metallic state. The theory explains
the large incoherent component of the optical conductivity
accompanied by a strong suppression of Drude weight observed
experimentally and is also consistent with measurements of the
specific heat. The fact that the anomalous transport properties in 
the ferromagnetic phase are described well supports the
orbital-liquid idea \cite{ISH97} and suggests that
the orbital degrees of freedom coupled to the lattice are of 
relevance in the metal-insulator crossover driven by the
magnetic transition in manganites.
This transition, which due to the double-exchange mechanism
is accompanied by a reduction of the mobility of charge carriers,
suppresses the energy scale ($\propto xt$) of orbital fluctuations 
and enhances the Jahn-Teller coupling\cite{MIL98} of these 
fluctuations to the lattice.
Apparently, the picture of an orbital liquid which is
quantum disordered by the metallic motion of holes 
is no longer valid above $T_c$; a disorder-order crossover
in the orbital sector is in fact indirectly indicated
by the temperature dependence of local lattice distortions 
experimentally observed. \cite{BIL96,BOO98}
We speculate this crossover to control the metal-insulator 
transition by further reducing the mobility of charge carriers, thus 
playing an essential role in enabling the formation of small lattice 
polarons and their localization above $T_c$.


We wish to thank P.\ Fulde and P.\ Horsch for valuable
discussions.



\begin{references}
\bibitem{COM1} The small upturn observed in the experimental
curve above $1.6 \text{ eV}$ is beyond the scope of the
present model describing the low-energy part of the spectrum
at $T=0$.
\end{references}
\end{document}